# Nano- and microscale apertures in metal films fabricated by colloidal lithography with perovskite nanocrystals

*Davide Spirito†, Javad Shamsi‡, Muhammad Imran, Quinten A. Akkerman, Liberato Manna, and Roman Krahne*

ᵃNanochemistry Department, Istituto Italiano di Tecnologia, Via Morego 30, 16163 Genova, Italy

ABSTRACT We demonstrate patterning of metal surfaces based on lift-off of perovskite nanocrystals that enables the fabrication of nanometer-size features without the use of resist-based nanolithography. The perovskite nanocrystals act as templates for defining the shape of the apertures in metal layers, and we exploit the variety of sizes and shapes that can be controlled in the colloidal synthesis to demonstrate the fabrication of nanoholes, nanogaps and guides with size smaller than the wavelength of light in the visible spectrum. The process can be readily integrated with standard lithography and etching techniques for the creation of more complex structures.

**Introduction**

Colloidal lithography has emerged as a viable low-cost alternative to electron-beam lithography (EBL) for the fabrication of nanostructures[1–3]. Its usual implementation employs micro- and nanospheres, made of polymers (e.g., polystyrene, poly(methyl methacrylate)) or inorganic materials (e.g., silica). After deposition on a target substrate, further processing (metal deposition, wet or dry etching) results in the formation of a variety of shapes on the substrate: circles, triangles, disks, rings, as well as 3D structures such as pillars and bowls. The resulting nanostructures and patterns have demonstrated peculiar optical [4], plasmonic [5,6], and magnetic [7] properties, and such processing can modify surface properties such as wettability. These features have found application in many different fields, for example in optics, energy harvesting[8] and storage[9], sensors[10], and biology[11]. However, the set of shapes that can be obtained from micro/nanospheres is limited by the circular projection of a sphere onto a plane, or by the projection of their packed arrays, and a more complex, multiple-step fabrication is required for more elaborate patterns.[12] Expanding the library of shapes that can be obtained in a simple, one-step fabrication process would strongly enhance the versatility of colloidal lithography. For example, shapes such as long guides[13–15], or rectangular slots, or gaps in metal conductors[16] are not possible with spheres, yet these features would be highly appealing for nanophotonic or plasmonic applications,[17] electronics, and nanofluidics.[18]

Here, we use perovskite nanocrystals as templates for different patterns. The colloidal synthesis of these nanocrystals is quite advanced and allows for control of their shape and size.[19–21] We demonstrate the patterning of nanometer-size features in metal layers with a variety of configurations that depend on the size and shape of initial template: nanoholes from nanocubes and platelets, nanogaps and nanoguides from nanowires. Furthermore, we combine the colloidal lithography with the deposition of luminescent quantum dots, and show how the process can be combined with standard EBL lithography, thus enabling the fabrication of more complex structures.

Figure 1 illustrates the straightforward fabrication process that consist of drop-casting of the nanocrystals, metal deposition and lift-off in a suitable solvent assisted by sonication.

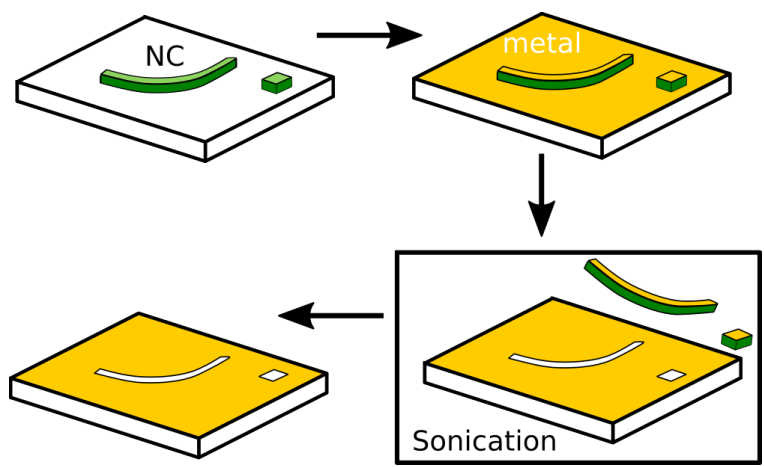

**Figure 1.** Illustration of the fabrication steps involved in the colloidal lithography process proposed in this work: First NC are deposited onto a substrate, then the metal layer is deposited, and finally the nanocrystals are removed by sonication assisted lift-off, which results in shape-controlled apertures in the metal film.

**Results and Discussion**

We employed nanocrystals of different shapes from the family of organic-lead halide perovskites, in particular $CsPbBr_3$ nanocubes[19], $CsPbBr_3$ nanowires[22], and $CsPbBr_3$ and $CsPbI_3$ nanosheets[21]. Figure S1 in the Supporting Information shows transmission-electron-microscopy (TEM) images of nanocubes and nanowires used in these experiments. In a typical colloidal lithography process, we deposit diluted colloidal solutions of the NCs (dispersed in toluene) to obtain individual nanocrystal structures on the surface. SEM inspection of the substrates revealed that drop-casting of diluted NC dispersions led to population of individual NC and small NC clusters. With an additional rinsing step (in hexane or isopropanol) it was possible to remove the larger NC clusters from the surface, which also cleaned the surface from other residues.

After NC deposition, the sample is coated with a metal layer by electron-beam evaporation, followed by a lift-off in acetone, chloroform, methanol, or a TMAH-based solution (ma-D 533 from Micro Resist Technology GmbH). The lift-off can take from several minutes to hours, and can be assisted by ultrasonication, but typically we obtained the best results with duration of 15-30 minutes under sonication. This colloidal lithography process leads to a metal film with holes, whose size and shape are determined by the NCs. Figure 2 shows scanning electron-beam microscopy (SEM) images of the different apertures in the metal film that we obtained: a square hole with few tens of nm edge length defined by $CsPbBr_3$ nanocubes (see Figure S2 for a statistical analysis), a hexagonal aperture with around 2.2 μm diameter by perovskite platelets, different rectangular shapes that are produced by nanosheets, and stripe-like apertures that result from nanowires. Nanowires are particularly appealing as templates because they can lead to straight or curved apertures, to crossings, and to loops.[23] A statistical analysis on the looped waveguides like the one depicted in Figure 2f is presented in Figure S3.

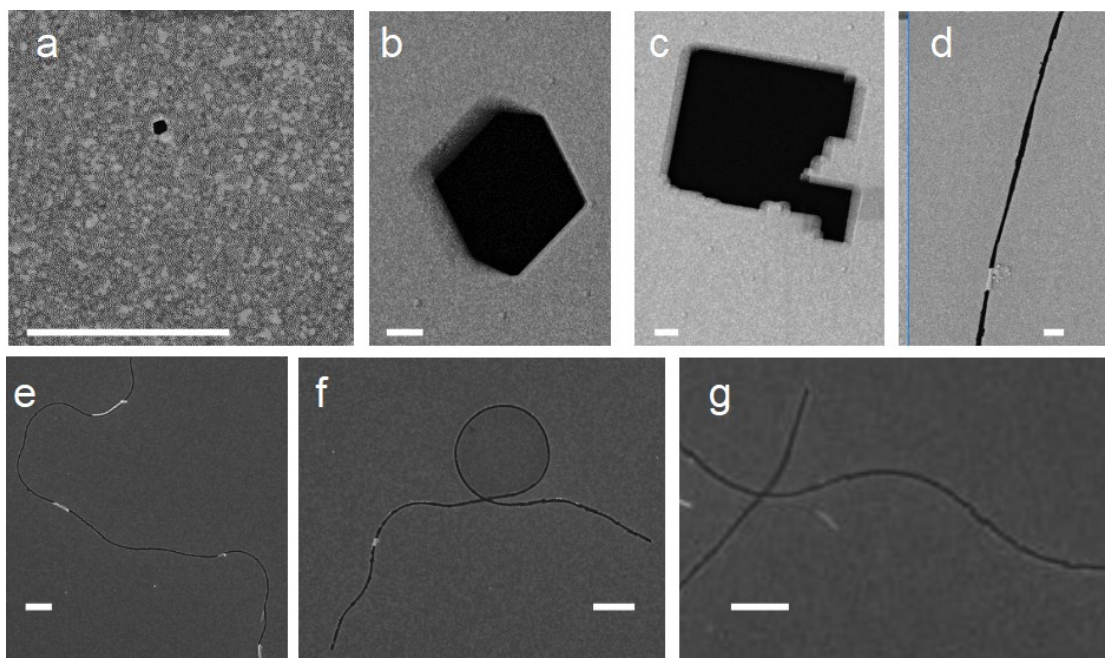

**Figure 2.** SEM images of the different kinds of apertures that were obtained by colloidal lithography using CsPbBr$_3$ nanocrystals as templates: (a) nanosized rectangle, (b) submicron hexagon, (c) micron-size combination of rectangles, (d) straight line with few tens of nm width, (e) curved line, f) loops, (g) line crossing. Scale bar is 500 nm in all images. The metal layer consisted on Ti/Au with the following thickness: (a) 3/10 nm, (b,c) 4/40 nm, (d) 3/30 nm, (e) 3/10 nm, (f,g) 5/20 nm.

Instead of CsPbBr$_3$ perovskite nanowires, we also prepared PbBrOH nanowires by simply heating up PbBr$_2$ in oleyamine at 200 °C for one hour in air. Such PbBrOH nanowires should be robust to organic solvents, while perovskite nanowires are known to dissolve in acetone, chloroform, methanol or isopropanol.[24] We tested the lift-off process using both nanowires made of PbBrOH, and from CsPbBr$_3$, to assess if our colloidal lithography process is due to mechanical lifting of the NC, or caused by a dissolution of the NC itself. As shown in Figure S4, we did not observe any lift-off with the PbBrOH wires, even for extended sonication up to 90 min.

On the contrary, the CsPbBr$_3$ perovskite nanowires were removed after 5 min of sonication. Therefore, we conclude that the perovskite NCs are dissolved by the solvent, and the metal layer is removed in a similar fashion as in resist-based lithography.

Since nanoscale apertures are particularly appealing in plasmonics, we explored different plasmonic metals for the process, namely Au and Ag. In both cases we used a thin (few nm) Ti layer for better adhesion of the metal layer to the substrate. In principle, the lift-off works for both Au and Ag, however, while for Au the apertures reflect precisely the nanocrystal shape, for Ag the edges appear corrugated and consequently the obtained aperture is larger and its shape is less defined (Figure S5). Cracks in CsPbI$_3$ sheets coated with Ti/Ag further support that the Ag in the metal layer reacts with the perovskite nanocrystal.

Nanosize holes in metal films are an interesting system with appealing optical properties[6,25–27]. To this end, the most promising NC templates in our approach are CsPbBr$_3$ nanocubes that have an edge length of ~14 nm (see Supplementary Fig. S1). We deposited these NCs from diluted dispersions in toluene by drop-casting, and applied the additional rinsing step in hexane or isopropanol to remove larger clusters of NCs from the surface. After metal deposition, we obtained holes with a diameter of around 22 nm that have a relatively narrow size distribution, as can be observed in Figure 3 a,b. Without the washing step, the holes were defined by the larger NC clusters, which resulted in diameters around 60 nm and a broader size distribution (Figure 3c,d). We also tested NC deposition by spin-coating, but this typically led to areas with closely packed NCs that did not work well for the lift-off process (see Figure S6).

Nanoholes in metal films can be used to concentrate analytes exploiting the different hydrophobicity of the metal and the substrate, and nanoscale apertures in a metal films can act as an optical cavity[28]. We spin-coated a solution of CdSe/ZnS core-shell quantum dots with 3.4 nm diameter and emission centered at a wavelength of 560 nm (purchased from Sigma Aldrich, (Lumidot™ CdSe/ZnS 560) onto a metal film with apertures that were defined by clusters of perovskite nanocubes, leading to holes of around 60 nm diameter. Confocal fluorescence microscopy imaging (Figure 3e) revealed a fluorescence signal in the green spectral band from the holes, confirming the positioning of the luminescent quantum dots into the apertures of the metal film. Figure S7 shows a relatively large aperture in the metal film where the PL intensity is concentrated in the center.

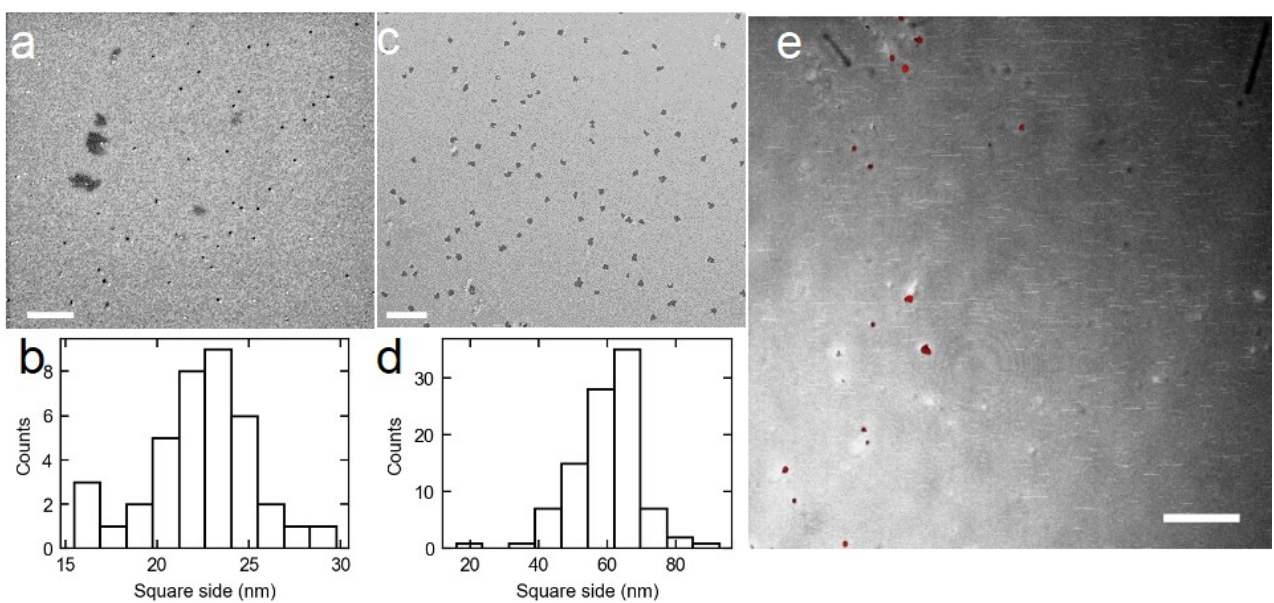

**Figure 3**. (a,c) SEM images of Ti/Au metal films films with nano-apertures produced by mostly single NCs (a), and NC clusters (c). Scale bar is 500 nm, and the metal layer was Ti/Au with 3/10 nm thickness. The corresponding size distribution of the edge length of the apertures is given in (b,d) and Figure S2b,c, respectively. (e) Confocal microscope image from a metal film with nanoholes after spin-coating with a solution of CdSe/ZnS quantum-dots emitting at 560 nm. The image shows an overlay of the reflectance of the surface (in grey scale) with the CdSe QD luminescence that appears in red color. Scale bar is 10 μm.

The colloidal lithography can be readily integrated with electron- or photoresist-based lithography to combine the apertures defined by the colloidal lithography with deterministic patterning of the metal film. One interesting combination is to use the colloidal lithography with nanowire templates to create nanosized gaps in electrodes that were defined by optical or electron-beam lithography. To this aim, we first spin the resist and define the desired pattern (i.e. the electrode in this case) by exposure and resist development. Then, a solution of the nanowires (in hexane) is drop-cast onto the PMMA/substrate sample, followed by metal deposition and lift-off in acetone. A second lift-off step in TMAH-based solution is performed to efficiently remove the nanowires. This process scheme is sketched in Figure 4a. Figure 4b shows an Au electrode with a gap of 70 nm that was obtained by the combination of electron-beam and colloidal lithography. We note that the fabrication of such narrow gaps in electrodes with large width (exceeding 1 μm) is highly challenging in the fabrication be EBL alone.

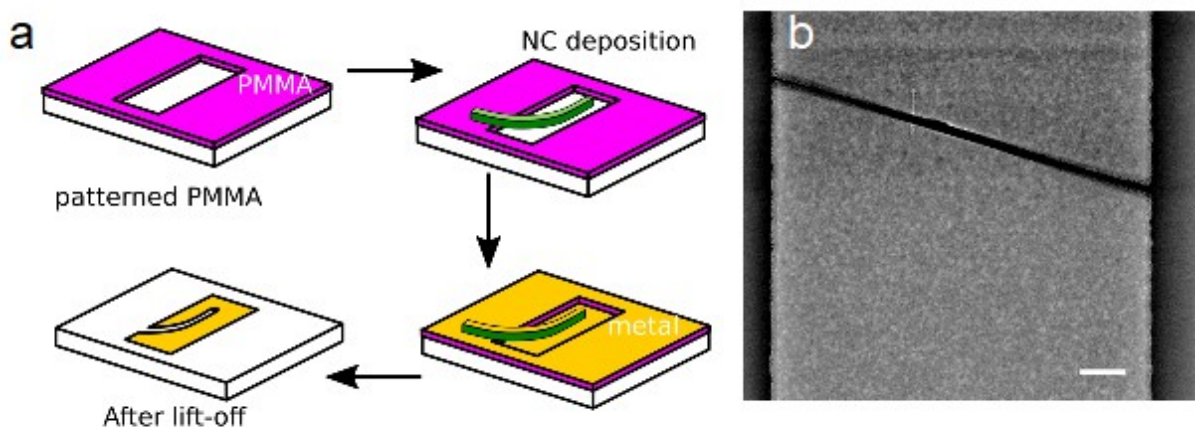

**Figure 4.** (a) Schematic illustration of the combination of conventional lithography with colloidal lithography. (b) SEM image of a nanogap in an Au electrode, where the gap was produced by colloidal lithography, and the electrode was defined by EBL. Scale bar is 250 nm, and the metal layer is Ti/Au with 3/10 nm thickness.

**Conclusions**

We demonstrated how perovskite nanocrystals can be used as templates for the creation of nanoscale apertures in metal films. This functionality is driven by the dissolution of the perovskite material during the lift-off process in organic solvents. The diversity of the possible shapes of the perovskite nanocrystals enables the fabrication of a variety of aperture shapes such as rectangles, hexagons, and straight and bend lines as well as loops. Such complex apertures in metallic films are highly appealing for far- and near-field optical studies. Further improvement of our colloidal lithography process could be achieved by elaborate self-ordering of the template NCs, or by controlled deposition, for example by electrophoresis. Combination with other processes such as standard resist-based lithography and dry-etching can pave the way to more elaborate 3D structures.

**Experimental section:**

<u>Lift-off without PMMA patterning:</u> Substrates (Si, Si/SiO2, glass) were cleaned in an ultrasonic bath with acetone, followed by isopropanol, and then dried under nitrogen flow. The colloidal solution of the NCs (typically 10 µl of toluene dispersion) was dropcast onto the substrate, and optionally followed by a rinsing step (in hexane or isopropanol) to remove larger NC clusters or other residues. Metals (Ti, Au, Ag) were evaporated in a Kenosistec e-beam evaporator with a deposition rate 0.3 Å s$^{-1}$ at a base pressure 10$^{-6}$ mbar.

<u>Lift-off with PMMA patterning:</u> For experiments with deposition of NCs on pre-patterned PMMA-coated substrates, we used Si/SiO$_2$ samples, cleaned as described above. PMMA was spin-coated onto them, and the pattern was exposed with a Raith 150-II EBL system. After the resist development, the NCs dispersed in hexane were drop-cast onto the substrate. We choose hexane as solvent as it does not degrade the PMMA. Then the metal layer is deposited as described above, and a first lift-off in acetone (ultrasound-assisted) is performed to remove the PMMA. However, this step is not sufficient to remove the NCs, and therefore a second lift-off step in TMAH-based solution is applied.

<u>Metal film imaging</u>: SEM images were acquired with the Raith 150-II system used for EBL or with a FEI Helios NanoLab DualBeam 650. Analysis of the SEM images for the determination of the nanohole size was performed with Gwyddion.

Corresponding Author: roman.krahne@iit.it

Present Addresses

†IHP – Leibniz-Institut für innovative Mikroelektronik, Im Technologiepark 25, 15236 Frankfurt (Oder), Germany

‡ Cavendish Laboratory, Department of Physics, University of Cambridge, JJ Thomson Avenue, Cambridge CB3 0HE

Author Contributions

The manuscript was written through contributions of all authors. All authors have given approval to the final version of the manuscript. ‡These authors contributed equally. (match statement to author names with a symbol)

ACKNOWLEDGMENT

This work was funded by the EU Horizon2020 MSCA Rise project "COMPASS-691185" and by the seventh European Community Framework Programme under Grant Agreement No. 614897 (ERC Consolidator Grant "TRANS-NANO").